\documentclass[reprint,pra,superscriptaddress,twocolumn]{revtex4}
\usepackage{units}
\usepackage{amsmath}
\usepackage{amssymb}
\usepackage{graphicx}
\usepackage{bm}
\usepackage{multirow,microtype,color,relsize,ulem}

\newcommand{\be}{\begin{equation}}
\newcommand{\ee}{\end{equation}}
\newcommand{\Ahat}{\hat{H}}
\newcommand{\Ehat}{\hat{E}}

\newcommand{\PT}{$\cal PT$}
\newcommand{\nr}{n_R}
\renewcommand{\ni}{n_I}
\newcommand{\dz}[1]{\frac{d #1}{d z}}
\newcommand{\ddz}[1]{\frac{d^2 #1}{d z^2}}

\newcommand{\ddx}[1]{\frac{\partial^2 #1}{\partial x^2}}
\newcommand{\ddy}[1]{\frac{\partial^2 #1}{\partial y^2}}
\newcommand{\re}[1]{\text{Re}[{#1}]}
\newcommand{\im}[1]{\text{Im}[{#1}]}
\newcommand{\Eq}[1]{Eq.~(\ref{#1})}
\newcommand{\Fig}[1]{Fig.~\ref{#1}}

\newcommand{\inner}[2]{\langle {#1}|{#2} \rangle}

\begin{document}
\title{Anomalous transient amplification of waves in non-normal photonic media}

\author{K. G.~Makris}
\affiliation{Department of Electrical Engineering, Princeton University, Princeton, New Jersey 08544, USA}

\author{L. Ge}
\affiliation{Department of Electrical Engineering, Princeton University, Princeton, New Jersey 08544, USA}
\affiliation{\textls[-38]{Department of Engineering Science and Physics, College of Staten Island, CUNY, New York 10314, USA}}

\author{H. E. T\"ureci}
\email{tureci@princeton.edu}
\affiliation{Department of Electrical Engineering, Princeton University, Princeton, New Jersey 08544, USA}

\begin{abstract}
Dissipation is a ubiquitous phenomenon in dynamical systems encountered in nature because no finite system is fully isolated from its environment. In optical systems, a key challenge facing any technological application has traditionally been the mitigation of optical losses. Recent work has shown that a new class of optical materials that consist of a precisely balanced distribution of loss and gain can be exploited to engineer novel functionalities for propagating and filtering electromagnetic radiation. Here we show a generic property of optical systems that feature an unbalanced distribution of loss and gain, described by non-normal operators, namely that an overall lossy optical system can transiently amplify certain input signals by several orders of magnitude. We present a mathematical framework to analyze the dynamics of wave propagation in media with an arbitrary distribution of loss and gain and construct the initial conditions to engineer such non-normal power amplifiers. Our results point to a new design space for engineered optical systems employed in photonics and quantum optics.
\end{abstract}

\date{\today}

\maketitle

\section{Introduction}
\label{sec:intro}

In practice, dynamical systems are never completely isolated and interact with their environment. In quantum and wave systems this interaction appears as dissipation of energy and other system properties. The conventional route to fighting dissipation relies on either the minimization of coupling to the environment or an external energy source that replenishes the energy or information lost. In optical systems this is accomplished by engineering the medium in which radiation propagates and through an external drive, respectively. Similarly, in engineered quantum optical systems such as superconducting circuits \cite{Girvin1}, recent interest is revolving around preservation of coherence and entanglement by engineering the coupling to a dissipative environment \cite{Girvin2}, an approach referred to as quantum bath engineering \cite{bath}. In optical systems, manipulation of the spatial distribution of the real part of the refractive index is today the cornerstone of modern photonics \cite{PhC,real}. However, only recently are the implications of engineering the imaginary part of the refractive index - that represents gain or loss - under scrutiny in nano-structured complex photonic structures (e.g. \cite{ring}) that in addition can feature electrical tunability (\cite{Liertzer2}).

Much of this early work on loss engineering has focused on optical configurations where the distributed loss and gain are in perfect balance. Referred to as Parity-Time (\PT) symmetric \cite{Bender1,Bender2} photonic systems \cite{PT_theory,PT_exp}, because the equations are isomorphic to quantum mechanics with a non-Hermitian Hamiltonian that is invariant under combined operations of spatial parity (${\cal P}$) and time-reversal (${\cal T}$), these optical systems in addition rely on a perfect spatial symmetry of the structure. Such novel structures have been experimentally realized with optical waveguides and fiber networks \cite{PT_exp} as well as with micro cavity lasers \cite{ring} and represent a class of optical systems where the deliberate introduction of loss and its spatial distribution along with gain can achieve new functionalities with potential applications as optical isolators and switches \cite{isolator}. Parity-time symmetric systems in the context of metamaterials \cite{PT_meta} and active plasmonics \cite{PT_plasmo} have also attracted considerable attention over the last years.

Recent experimental work \cite{PT_exp,ring} however has shown that in most photonic applications it is difficult to implement designs in which optical loss and gain are perfectly balanced. Furthermore in a number of optical systems e.g. active plasmonic structures, the overall loss generally dominates the optical gain. For the full realization of the potential of "loss engineering", it is of interest to examine the properties of the larger class of systems that we refer to as "non-normal optical potentials". These feature arbitrary spatial distributions of gain and loss not subject to any spatial symmetry requirements. Are there any generic properties of such structures that may present an outstanding promise for novel applications? This is the question we attempt to answer here.

We study wave-propagation in multimode optical waveguides that feature distribution of gain and loss that is not balanced, in particular one where {\it the optical loss dominates}. As a direct outcome, all eigenmodes of the system are decaying with propagation distance, whether they are confined by the optical potential (bound modes) or not (radiation modes). We show that depending on the initial conditions, the injected power can nevertheless be amplified, in some cases by several orders of magnitude. We find that the initial field distributions that gives rise to the maximum growth is {\it not} localized only within the gain regions, contrary to what one would expect. We subsequently show that the appropriate initial field that leads to the maximum possible power growth can be calculated using singular vectors of the {\it propagator} of the system. This direction is largely unexplored in the non-normal optics literature \cite{Cheng,Siegman1,Siegman2,Berry1,lasers1,lasers2,Liertzer,Liertzer2,Snyder,Konotop} and can have potential applications for transient power amplification and directed energy transfer in inhomogeneous media.

\section{Photonic structures as non-normal dynamical systems}
\label{sec:general}

We consider the propagation of optical waves in spatially complex photonic structures with a preferred axis for propagation ($z$), characterized by a complex index of refraction that is on average loss-dominated. Under the paraxial approximation, the dynamics of the slowing-varying field amplitude $\Phi(x,y,z)$ is captured by a Schr\"{o}dinger-like equation (paraxial equation of diffraction) $\frac{\partial \Phi}{\partial z}=\Ahat\Phi$, where $\Ahat$ is the evolution operator of the system. This analogy with $z$ playing the role of time $t$, has been used extensively to map quantum mechanics to the optical domain and is the basis for a number of photonic quantum simulation schemes \cite{real,aspuru}. Since the refractive index we consider is complex, the operator $\Ahat$ is generally non-normal \cite{TrefethenBook} i.e. it does not commute with its adjoint $[\Ahat,\Ahat^\dagger]\neq0$.

Assuming translational invariance along the propagation direction $z$, our understanding of the beam dynamics is typically based on the spectrum of $\Ahat$. In a non-normal system, the eigenmodes $\phi_n(x,y)$ of $\Ahat$ are non-orthogonal and the eigenvalues $\lambda_n = \gamma_n + i\beta_n$ complex. Using the biorthogonality relationship between the eigenmodes $\phi_n(x,y)$ of $\Ahat$ and the associated adjoint eigenmodes $\tilde{\phi}_m(x,y)$, namely   $\inner{\phi_m}{{\phi}_n} \equiv \int_{-\infty}^{\infty} \,dx\,dy\,\tilde{\phi}_m^*(x,y)\phi_n(x,y) = \delta_{mn}$, the dynamics of any arbitrary field can be written as a superposition: $\Phi = \sum_{n=1}^\infty c_n\phi_n(x,y) e^{i\beta_nz}e^{\gamma_nz}$. This expansion is of course valid provided that the set of the non-orthogonal eigenmodes is complete. The general approach we are going to take to investigate the transient growth dynamics however, as we shall show later, does not rely on such an expansion.

For a structure with the overall loss greater than the gain (to be defined below), the spectrum typically but not generally consists of only decaying eigenvalues (i.e. $\gamma_n<0$ in the convention chosen here), and such a system is described to be subject to "modal loss". In such situations, the input beam decays over a long propagation distance $z$. In view of the above, the answers to the fundamental questions related to a lossy amplifier (What is the maximum achievable amplification? What are the corresponding optimal initial conditions for maximum amplification at a given distance $z$? Can the system  exhibit gain in the asymptotic long-distance limit as well?), appear to be highly nontrivial and counterintuitive. The systematic examination of these fundamental questions regarding the transient amplification of decaying waves is the main focus of this paper.

\section{Wave propagation in non-normal optical waveguides}
\label{sec:wave}

\begin{figure}[tb]
\includegraphics[clip,width=1\linewidth]{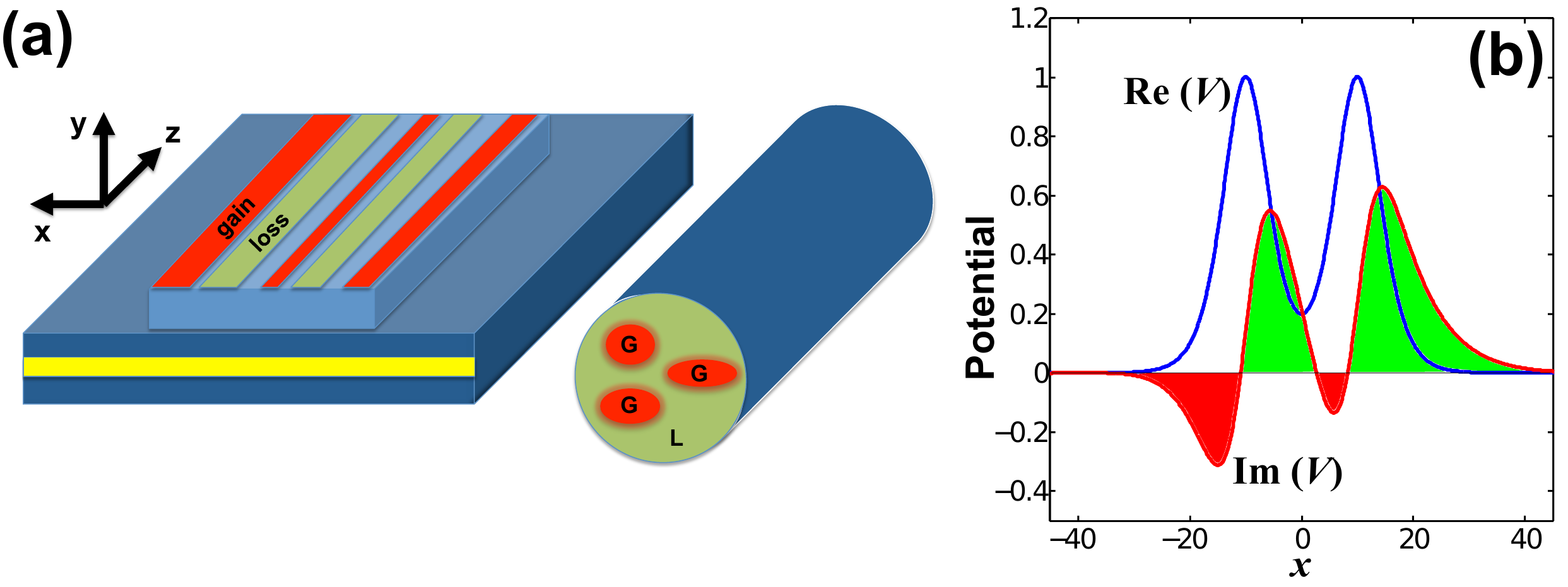}
\caption{Schematic depiction of non-normal photonic waveguide structures in (a) semiconductor slab waveguide and an optical fiber. In both figures the green and red shaded regions represent the lossy and gainy regions, respectively. The propagation distance is denoted with $z$. (b) Real (blue curve) and imaginary (red curve) part of the non-normal optical potential we consider in the text, for $g=0.5$. The green filled areas denote the lossy parts of the potential, while the red filled areas stand for the gain parts.}\label{fig:schematics}
\end{figure}

We begin our analysis by considering optical wave propagation in a generic non-normal potential. These potentials can be generated using either planar waveguides or optical fibers, like those depicted in \Fig{fig:schematics}(a). Such potentials, are on average lossy, and characterized by a complex index of refraction $n(\bm{r}) = n_0 + \delta n(\bm{r})$. Here $n_0$ is the real-valued background index guiding the propagating wave, and $\delta n$ is complex-valued refractive index spatial modulation (typically $\delta n \ll n_0$). The optical wave $\Psi(\bm{r},t)$ of frequency $\omega$ can be expressed as $\Psi=\Phi(x,y,z)e^{i(n_0k_0z-\omega t)}$ with $k_0\equiv\omega/c=2{\pi}/{\lambda_0}$ where $c$ is the speed of light in vacuum, $\lambda_0$ the optical wavelength and $\Phi(x,y,z)$ is the slowing-varying field amplitude in the propagation direction $z$. The normalized paraxial equation of diffraction for the electric field envelope takes the following form\cite{real}:
\begin{align}
i\frac{\partial \Phi}{\partial z} +\ddx{\Phi} +\ddy{\Phi} +V(x,y)\Phi=0,
\end{align}
The above equation can be written in the dynamical systems form $\frac{\partial \Phi}{\partial z}=\hat{H}\Phi$, where $\hat{H}$ is the evolution operator of the system is given by:
\begin{align}
\Ahat = i \left(\ddx{} + \ddy{} \right) + iV(x,y)
\end{align}

One can easily see the non-normal nature of $\Ahat$ for a complex optical potential $V \equiv \nr(x,y)+i g \ni(x,y)$. We choose $\ni$ to be positive (negative) for loss (gain). For convenience we introduce a dimensionless constant $g$ to adjust the relative magnitude of the imaginary part of the index to the real part {\it without modifying the loss-gain distribution}. For all cases, the overall imaginary part of the index is much smaller than the real part because $V$ is the index fluctuation on top of a large real background index $n_0$. The class of optical potentials that we investigate is characterized by a spatial average corresponding to net loss i.e. $\int_{-\infty}^{\infty} dx dy \, \ni (x,y) > 0$. This is what we meant by "overall loss" mentioned before. Note that this quantity is independent of $g$. Such inhomogeneous gain-loss landscapes do neither require any stringent spatial symmetries, nor can they be mapped to equivalent of \PT-symmetric optical potentials \cite{PT_theory} .

Unlike gain guiding in homogeneous potentials \cite{Siegman2}, where modal gain characterizes completely the dynamics and the power amplification of the optical fiber, our system has zero modal gain. In particular, even though there is material gain in the system, there is no modal gain since all the eigenvalue spectrum corresponds to decay eigenmodes, meaning modal loss. This class of lossy potentials can still be used as power amplifiers due to the physical existence of gain material. We are in particular interested in studying the evolution of the integrated intensity along the $z$ direction. In particular, we can derive the following expressions that describe the dynamics of power $P(z)\equiv\int_{-\infty}^{\infty} dx \, dy \, |\Phi(x,y,z)|^2$ along the $z$ direction. We can further derive the dependence of power on $z$ as  $\dz{P} = -2g\int_{-\infty}^{\infty} dx dy \, n_I |\Phi|^2 $ and $\ddz{P} = 4g\int_{-\infty}^{\infty}{\nabla n_I(x,y)}\cdot S_\perp(x,y,z)\,dx$, where $S_\perp(x,y,z)$ is the transverse component of the Poynting vector, defined as $S_\perp=\frac{i}{2}(\Phi \nabla \Phi^*-\Phi^* \nabla \Phi)$ . For the simple case of only loss (gain), we immediately find $\dz{P}<0(>0)$  for every value of the propagation distance $z$, indicating power decay or amplification. The situation is different for potentials that involve both gain and loss. In particular, we can understand from these relations that the power growth dynamics is not easily described, since we do not know a priori the diffraction evolution of a given input waveform and thus the sign of $\dz{P}$. At this point we have to add that the understanding of such maximal transient growth cannot rely on the existence of exceptional points \cite{PT_theory,PT_exp,PT_plasmo} or optical nonlinearities \cite{Snyder} used to achieved switching in coupled gain-loss structures. Instead, it can only be understood as an inherent characteristic of the underlying non normal dynamical system, in terms of pseudospectra and singular values.

\section{Decaying Eigenvalue spectrum of lossy amplifiers}
\label{sec:decay}

\begin{figure}[tb]
\includegraphics[clip,width=1\linewidth]{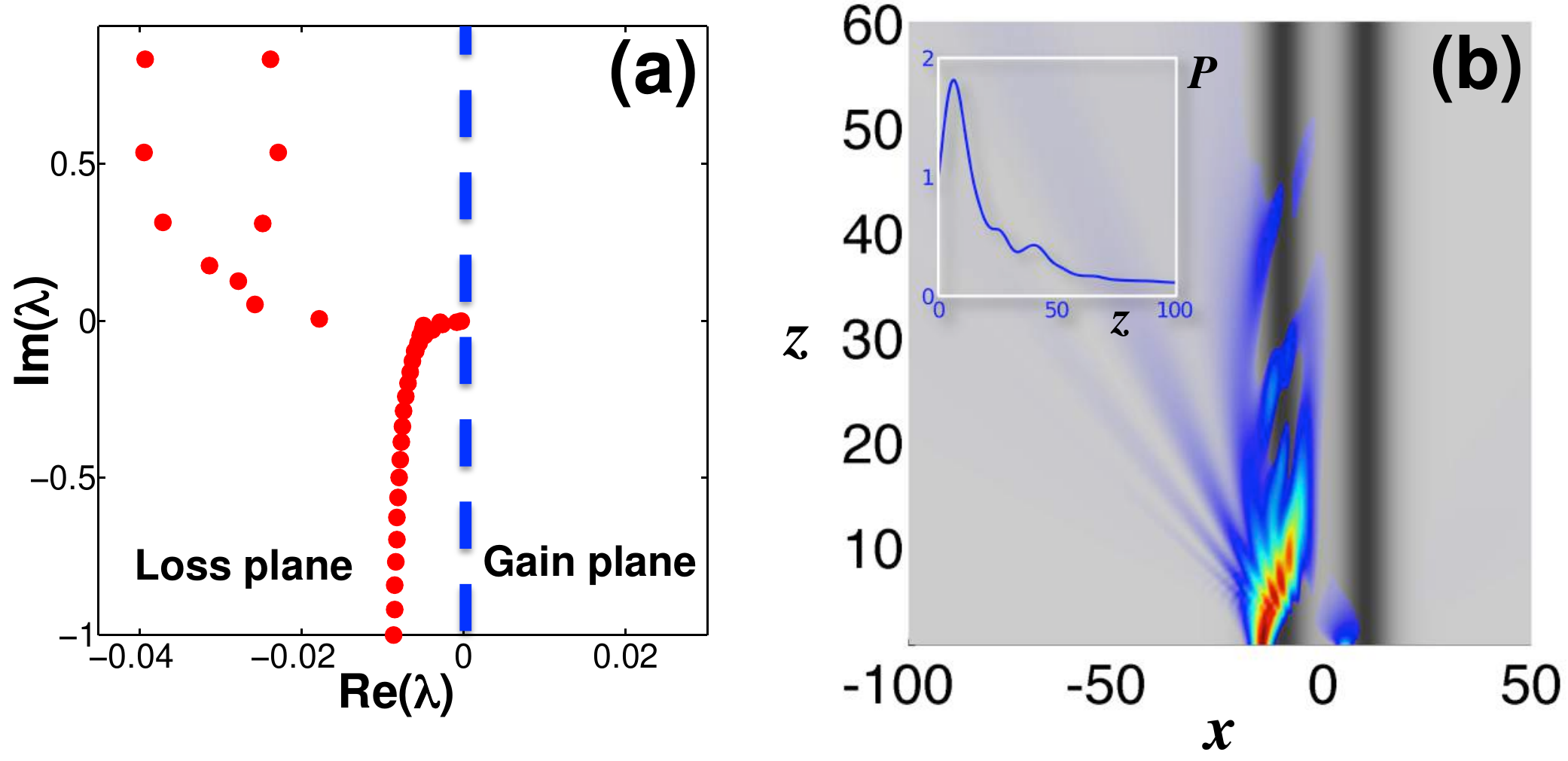}
\caption{ (a) Eigenvalue spectrum for $g=0.5$ plotted in the complex plane. (b) Intensity wave dynamics for the intuitive initial conditions. Namely the light is coupled only at the gain regions. For that initial condition, the inset depicts the optical power versus the propagation distance $z$. The two black regions represent the two waveguides and the inset the power versus the propagation distance $z$.}\label{fig:1d}
\end{figure}

Henceforth we exemplify our findings by using one-dimensional systems, and an example of a two-dimensional system is given in Appendix B. Before proceeding with the analysis of the power growth, is it crucial to first examine the nature of the eigenvalue spectrum of our lossy amplifier. Let us consider a specific example of non-normal multimode waveguide that is loss-dominated, like the one that is depicted in Fig.~\ref{fig:schematics}(b).  For propagation through such a waveguide ($g=0.5$ in this case) all  eigenvalues are located in the left half part of the complex plane (``loss plane"), as is shown in Fig.~\ref{fig:1d}(a). Even if the averaged $\ni$ is loss-dominated, since the system physically exhibits gain in some regions, one would expect that it can in principle amplify the propagating light. Basic theoretical considerations tell us however that in the large-$z$ limit, the total power will eventually always decay exponentially to zero for a waveguide characterized by modal loss (all eigenvalues in the loss plane). On the other hand, it is a basic fact of non normal operator theory \cite{TrefethenBook} that even if all the eigenvalues are in the loss plane, the system may exhibit non-exponential transient behavior \cite{TrefethenBook,growth1,growth2} before eventually entering the exponential decay regime. Indeed in Fig.~\ref{fig:1d}(b), we can see the beam dynamics versus the propagation distance $z$, {\it when the light is initially ($z=0$) coupled only to the gain regions}. The inset depicts the integrated intensity (optical power $P(z)$) over the propagation distance. Total power is initially amplified (by a factor of 1.5), but within a short distance, light diffracts out of the gain region and the total power then decays eventually to zero with $z$, as expected. We show below that coupling light to the gain regions is not the optimal initial condition to achieve maximum transient amplification.

\begin{figure}[tb]
\includegraphics[clip,width=0.8\linewidth]{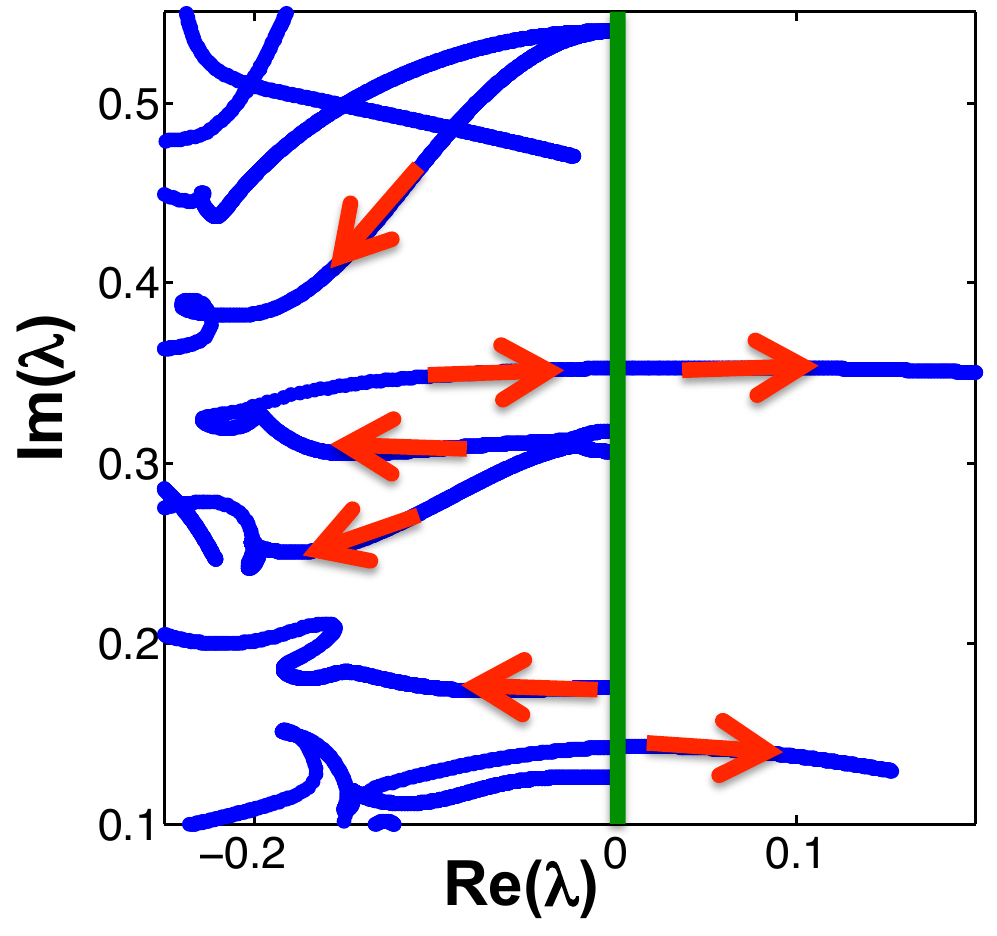}
\caption{Eigenvalue trajectories in the complex plane as $g$ increases from zero. At $g=0$ the system is hermitian and the eigenvalues are real. The red arrows denote the motion of the eigenvalues in the complex plane as the gain-loss amplitude $g$ varies.} \label{fig:path}
\end{figure}

Another important aspect is the understanding of the eigenvalue dependence on the only free parameter which is the amplitude $g$ of the imaginary part of the optical potential. As we increase its value from zero, we see that after some critical threshold ($g_c=3.65$) some eigenvalues cross the imaginary axis of the complex plane and experience gain, as is depicted in Fig.~\ref{fig:path}. For $g>g_c$ the system displays a transition to a globally amplifying behavior, i.e. in the asymptotic $z \rightarrow \infty$ limit, the power increases exponentially. This is found to be consistent with an eigenvalue of the system crossing over to the gain plane. This resembles the mathematical "phase transition" that characterizes all \PT-symmetric Hamiltonians \cite{Bender1,Bender2,PT_theory,PT_exp}. The difference is that we do not encounter exceptional points here. We emphasize that the system is still characterized by an overall loss, for, as pointed out before, that is independent of $g$.

	In view of the above, since all eigenvalues correspond to bound and radiation decaying eigenmodes, one would naturally expect that for every input waveform the optical power must decay monotonically with the propagation distance $z$, based on our intuition. That is not true however, and we find that transient growth of the optical power for certain input waveforms is possible. Therefore the eigenvalue spectrum fails to capture the transient dynamics intuitively, and a result we have to employ a different approach that illuminates the relationship between the latter and the non-normal nature of the system. Such approach is based on the notion of singular values of the propagator and the pseudospectrum \cite{TrefethenBook} of the non-normal evolution operator. We believe that our methodology may be applicable in the evolution dynamics of systems that are described by non-hermitian random matrices as well. This is for instance relevant in the case of Fokker-Planck equations \cite{FP}, where our analysis can be helpful in understanding the time dynamics of diffusion processes.

\section{Transient Power Growth and Singular Value decomposition}
\label{sec:SVD}

Before we establish the mathematical framework to analyze the transient behavior \cite{growth1,growth2}, it is crucial to define the meaning of the central quantity of interest in our study, the power amplification ratio for a given propagation distance $z$:
\begin{align}
G(z)\equiv \frac{P(z)}{P(0)} =\frac{\|\Phi(x,z)\|^2}{\|\Phi(x,0)\|^2} \label{eq:G}
\end{align}
where $\|f\|$ is the usual Euclidean norm of a function, $\|f\|^2\equiv\int_{-\infty}^{\infty} dx \, |f(x)|^2$. The $G(z)$ depends strongly on the input waveform. We will first discuss its upper bound at a given $z$ for all possible input waveforms:
\be
G_{\max}(z) = \sup \frac{\|\Phi(x,z)\|^2}{\|\Phi(x,0)\|^2}=\|e^{z\Ahat}\|^2  \label{eq:Gmax}
\ee
where the right hand side is the square of the matrix norm of the propagator of the system $e^{z\Ahat}$. At this point we have to note the conceptual difference of the growth ratios described by \Eq{eq:G} and \Eq{eq:Gmax}. In the first case $G$ is the ratio of the output over input power for a specific initial condition. On the other hand, \Eq{eq:Gmax} describes the maximum amplification ratio for all possible initial conditions at $z=0$. As such, we are first interested in finding estimates for $G_{\max}(z)$ as a preliminary characterization of our lossy amplifier.

As opposed to Hermitian or normal systems for which the spectrum of $\Ahat$ characterizes the entire dynamics, it's the {\it pseudospectrum} of $\Ahat$ that is the relevant construction for these estimates.  In particular, the lower and upper bounds of $G_{\max}(z)$ can be estimated using functional analysis theorems of non normal operators \cite{TrefethenBook} and are directly associated with the extension of the pseudospectrum cloud to the right half complex plane (Appendix~\ref{appA}). The $\epsilon-$pseudospectrum $\sigma_\epsilon(\Ahat)$ of a non-normal operator $\Ahat$ is defined by the union of all its eigenvalues in the complex plane, when subjected to all possible system perturbations below a certain magnitude. Practically speaking, the pseudospectrum is a pattern in the complex plane that shows how sensitive the eigenvalue spectrum $\sigma(\Ahat)$ is to random perturbations. For hermitian operators the spectrum and the pseudospectrum are almost identical. But for non-normal operators the two patterns can differ significantly, depending on the degree of the non-orthogonality of the corresponding eigenmodes.

\begin{figure}[tb]
\includegraphics[clip,width=1\linewidth]{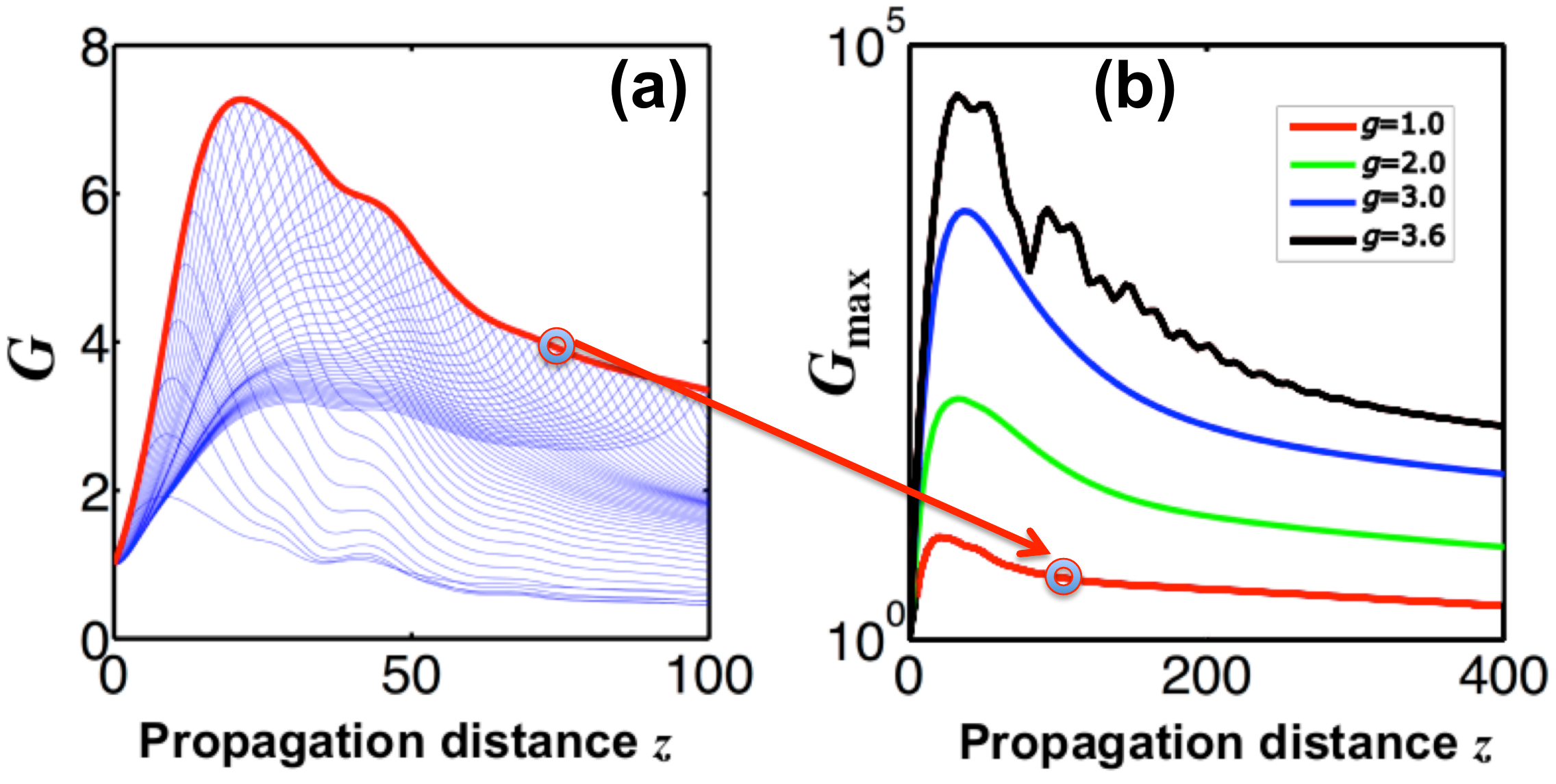}
\caption{(a) Power amplification for $g=1$ versus the propagation distance, for many different optimal initial conditions (thin blue curves). (b) Transient power growth in logarithmic scale as a function of the propagation distance $z$ for different values of the gain and loss amplitude $g$. The system is the same as the one considered in Fig.1. Note that the envelope of many different $G$ curves produces one $G_{max}$ curve.} \label{fig:three}
\end{figure}

The pseudospectrum analysis presented in the Appendix~\ref{appA} is very useful since it provides a geometrical method which can be used to estimate the magnitude of the maximum transient growth in a non-normal optical potential. However it does not provide a direct insight into {\it which} initial conditions lead to maximum amplification. We present below a method based on the singular value decomposition of the non-normal propagator $\hat{G}(z) = e^{z\Ahat}$ to construct these special initial conditions for a given structure. It can be shown that
\be
G_{\max}(z) = \|\hat{G}\|^2=\max (\sigma_n)^2,\label{eq:Gmax_SVD}
\ee
in which we have used the well known property that the matrix norm of an operator is given by its maximum singular value \cite{growth1,growth2}.  We note that the singular values $\sigma_n$ are real and non-negative even for a non-hermitian operator. Based on the above analysis, we calculate (\Fig{fig:three}) the maximum growth as a function of $z$ and for different values of the gain and loss amplitude $g$. To further demonstrate the meaning of the maximum power growth curve $G_{\max}(z)$, we emphasize again that for each value of $z$, this maximum growth is achieved by using a {\it different} input waveform. If we plot the power dynamics versus propagation distance for each one of these optimal initial conditions that lead to maximum growth, their upper envelope should give $G_{\max}(z)$. In other words the upper envelope of many $G(z)$ curves (produced only by the optimal initial conditions for every $z$) is the $G_{\max}(z)$ curve. We note that the exact bounds found through this analysis (\Fig{fig:three}(a)) agrees very well with the pseudo-spectrum based estimates given in Appendix~\ref{appA}  for $g=1$, namely $2\leq G_{\max} \leq 12$. \Fig{fig:three}(b) shows that the peak amplification rate increases dramatically as $g$ approaches the critical value $g_c=3.65$.

Giant amplification in nonhermitian structures is a well studied effect in the context of \PT-literature \cite{PT_theory,PT_exp,Konotop,pete2}. Close to an exceptional point the interference between nonorthogonal eigenmodes leads to huge amplification ratios and power oscillations. The understanding though of the maximal amplification ratio and under what conditions it can be achieved in a generic non-normal potential (non-\PT and without any exceptional point) is still a direction largely unexplored in the field of nonhermitian photonics. The nontrivial initial conditions that lead to such maximal amplification for a given value of the propagation distance $z$, and their connection to the eigenvalue spectrum, is the subject of next section.

\section{Optimal Initial conditions and their transient dynamics}
\label{sec:init}

So far we have not analyzed the physical content of the input waveform that leads to the maximum transient growth at a given $z$. We show below that the initial conditions for maximum amplification are complicated, and most remarkably are not localized merely in the gain regions. The input waveform to achieve the maximum transient growth at $z$ is given by the corresponding right-singular vector $\nu_n(x)$, defined by
\be
\hat{G} \nu_n(x) = \sigma_n v_n(x),\quad
\hat{G}^\dagger v_n(x) = \sigma_n \nu_n(x),
\ee
where $v_n(x)$ are the left-singular vectors. We note that the singular values $\sigma_n$ are real and non-negative even for a non-hermitian operator. The conclusion (\ref{eq:Gmax_SVD}) can be shown by writing
\be
G(z) = \frac{\inner{\Phi(x,0)|\hat{G}^\dagger \hat{G}}{\Phi(x,0)}}{\inner{\Phi(x,0)}{\Phi(x,0)}},
\ee
which is maximized when $\Phi(x,0)$ is the eigenvector corresponding to the largest eigenvalue of the hermitian operator $\hat{G}^\dagger \hat{G}$, which is exactly the right-singular vector of $\hat{G}$ with the largest $\sigma_n$:
\be
\hat{G}^\dagger \hat{G} \nu_n(x) = \sigma_n \hat{G}^\dagger v_n(x) = \sigma_n^2\nu_n(x).
\ee
Intuitively speaking, one would expect that if the input waveform is coupled to the gain regions only, it probably will lead to the maximum power growth. This is however, not true as we show below.

\begin{figure}[tb]
\includegraphics[clip,width=1.0\linewidth]{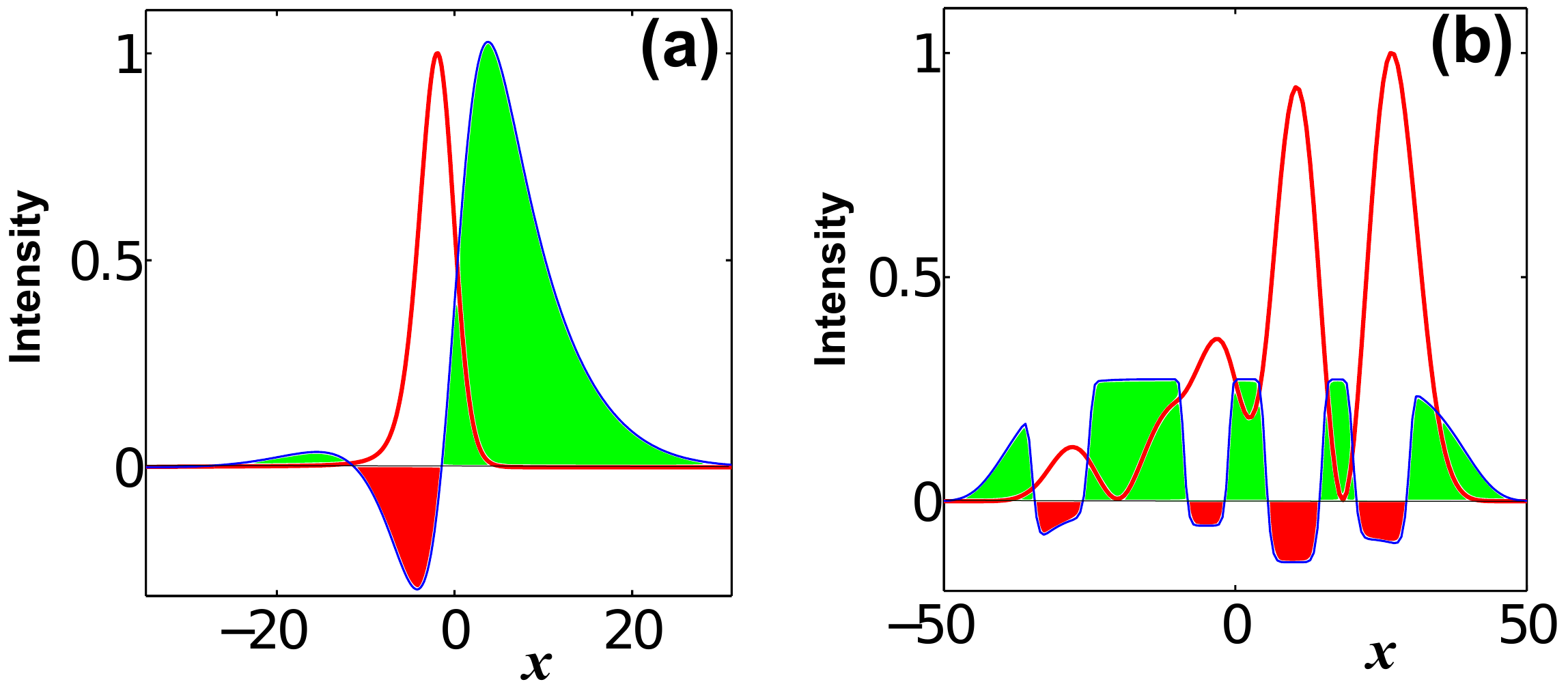}
\caption{Intensity plots (red lines) of the optimal initial conditions that lead to maximum transient growth for (a) a
single waveguide (b) a multimode waveguide with random distribution of its imaginary part. In both plots the blue
lines represent the imaginary part of the non-normal optical potential. }\label{fig:strongerLoss}
\end{figure}

To illustrate this important and counterintuitive result, we consider two other examples of non-normal potentials with net loss. The first one is that of a single waveguide with asymmetric gain-loss profile (Fig.~\ref{fig:strongerLoss}(a)) and the second one that of a multimode waveguide with random spatial distribution of gain and loss (Fig.~\ref{fig:strongerLoss}(b)). In both cases the optimal input waveform that achieves the maximum power growth (at $z=20$ and $z=180$, respectively) resides in both the gain and the loss regions, as we can see from its spatial distribution in Fig.~\ref{fig:strongerLoss}(a),(b). In all cases the common intuitive picture of coupling the input light only to the gain regions leads to a fast and small power growth and in any case not in the maximum achievable growth. On the other hand, the calculated initial field distributions lead to the maximum possible power growth for longer propagation distances. 

\begin{figure}[tb]
\includegraphics[clip,width=1\linewidth]{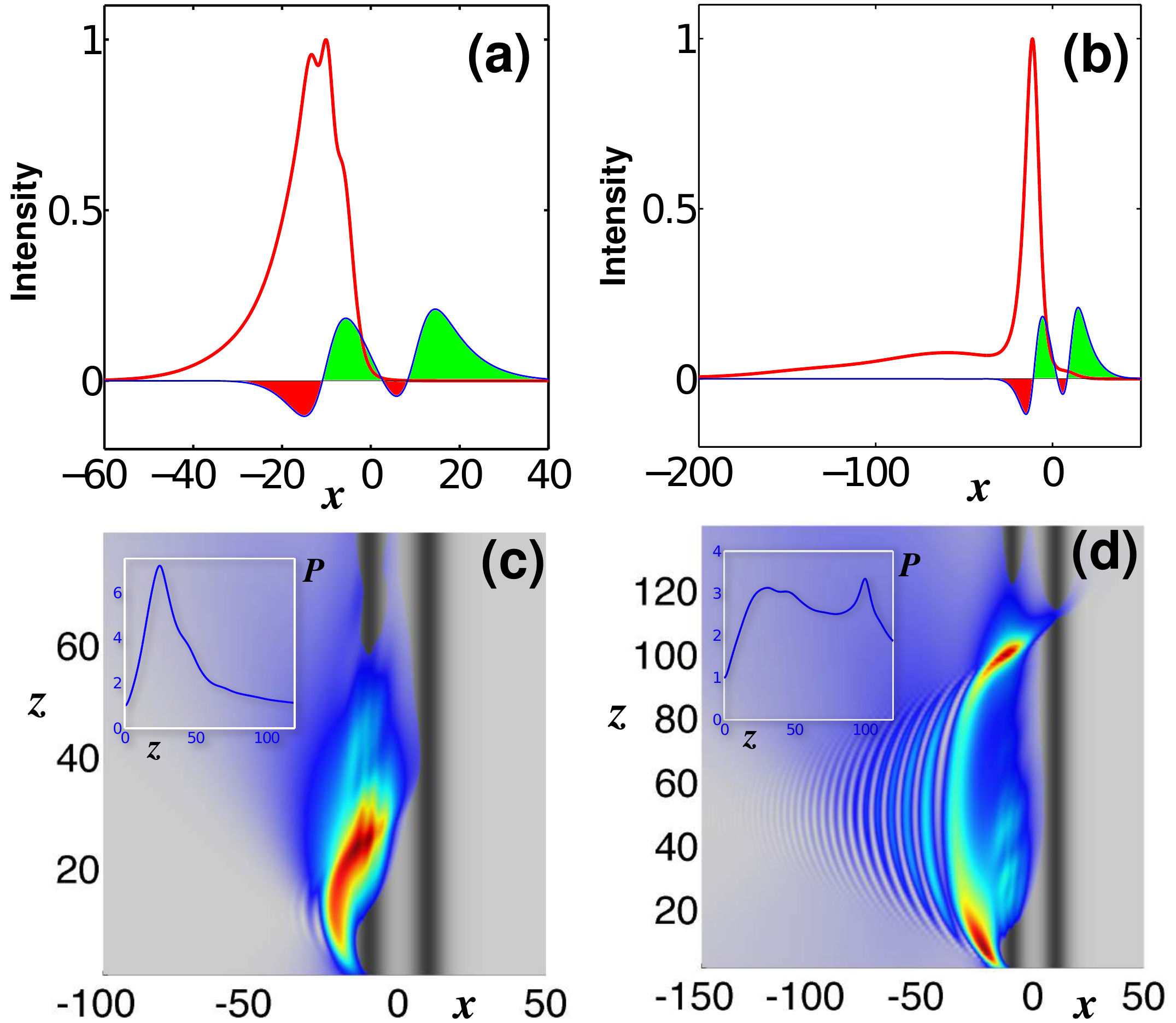}
\caption{(a),(b) Intensity profiles of two different inputs (thick red curves) along with the imaginary part of the refractive index for reference (thin blue line) (c,)(d) Beam diffraction dynamics of the corresponding input waveforms in (a) and (b). The two insets depict the optical power versus the propagation distance $z$ for the initial conditions given in (a),(b).}\label{fig:diff}
\end{figure}
Regarding our initial example of non normal multimode waveguide of Fig.~\ref{fig:schematics}(b), we analyze the optimal initial conditions for maximal growth in the following Fig.~\ref{fig:diff}). More specifically, to reach the corresponding maximum growth at $z=25$, a highly non-trivial input waveform is required, which has a significant overlap with the loss region (red curve in Fig.~\ref{fig:diff}(a)). Inspecting the power dynamics more closely, we find that such input indeed achieves the maximum achievable amplification ($G_{\max}(z=25)=7$), and it follows an interesting diffraction pattern (seen in Fig.~\ref{fig:diff}(c)) that has a prolonged overlap with the gain regions as it propagates in the $z$ direction, and it achieves the maximum possible growth at $z=25$, many wavelengths away from $z=0$. For an initial condition corresponding to peak amplification at a larger value of propagation distance $z=100$, (Fig.~\ref{fig:diff}(b)) the transient dynamics is illustrated in Fig.~\ref{fig:diff}(d). The beam follows an Airy-beam \cite{Airy} like diffraction pattern outside the waveguide region, in order to achieve, at the specific propagation distance, the maximum growth. This can be also viewed as a focusing effect that takes place under the confluence of gain/loss and diffraction.

In the previous paragraphs we showed the the input waveform that leads to the maximum transient growth at propagation distance $z$ is given by the right singular mode of the propagator $e^{z\hat{H}}$ that corresponds to the largest singular value, for a particular propagation distance $z$. So a natural question to ask is the relationship between the optimal waveform for maximum amplification and the eigenmodes of $\Ahat$. This can be studied by projecting the optimal initial condition $\Phi_{opt}(x,z=0)=\nu_n(x)$, that leads to the maximum power growth, onto the biorthogonal eigenbasis of $\hat{H}$ that consists of the set of eigenfunctions $\phi_n,\tilde{\phi}_n$. More specifically the superposition coefficients $c_n$ of the projected optical input beam $\Phi_{opt}$ to the biorthogonal basis in given by the following equation: $c_n= \inner{\phi_n}{{\Phi}_{opt}(x,z=0)} /   \inner{\phi_n}{{\phi}_n}$. This expression provides us with an alternative interpretation of the transient growth. We note at this point that all the projection coefficients considered here lead to convergent series expansions \cite{Siegman3}. For high enough values of the projection coefficients $c_n$ one expects to see transient power growth for small values of propagation distance $z$. In order for the coefficients $c_n$ to obtain high values, two conditions should be satisfied. The first condition is that the numerator has to have high values. In order for this to happen the initial condition $\Phi_{opt}$ must be nearly orthogonal with the adjoint eigenmodes $\tilde{\phi}_n$.  The second condition is related to the low values of the denominator, which is directly linked to the Peterman excess noise factor \cite{Berry1,pete1,pete2} $K_n=1/|\inner{\phi_n}{\phi_n}| ^2$. In laser physics, this factor determines the deviation of the quantum-limited laser linewidth of leaky cavities from its Schawlow-Townes value.
\begin{figure}[tb]
\includegraphics[clip,width=0.8\linewidth]{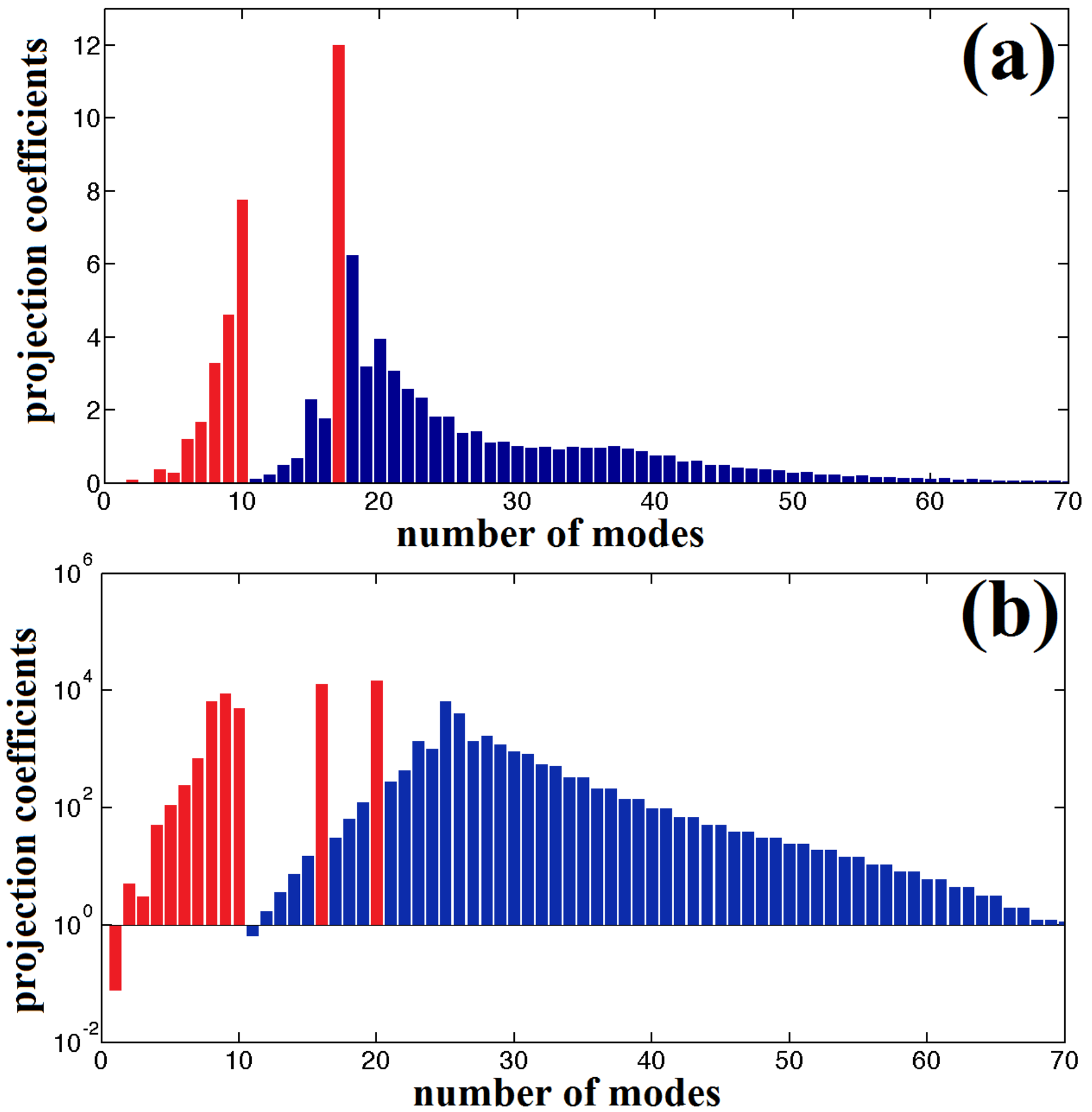}
\caption{Absolute value of projection coefficients versus the number of modes of the initial condition the achieves the maximum power growth at for (a) $z=25,g=1$ and (b) $z=38,g=3$. The second diagram is in logarithmic scale. The red bars correspond to bound modes whereas the blue ones to radiation modes of the eigenvalue spectrum. Also the modes are sorted based on their imaginary part of their corresponding eigenvalues.}\label{fig:peter}
\end{figure}
We have to note at this point that the distribution of the values of the Petermann factor for different modes follows the distribution of the absolute value of the projection coefficients. In Figs.~\ref{fig:peter}(a,b)  the absolute value of the projection coefficients $|{c_n}|$ is plotted versus the mode number for the optimal field inputs at $z=25, g=1$ and $z=38, g=3$.  The second figure is plotted in logarithmic scale and we can see which modes of the eigenspectrum contribute more to the superposition of the optimal input. We denote with red color the bound modes of the potential. We can see that the optimal initial beam is composed of both bound and radiation modes, but most of its energy is contained in the most non-orthogonal bound modes. In other words, the modes that contribute most to the pseudo spectrum cloud have also the highest projection coefficients and as a result contribute more to the transient growth.  We can also understand that the eigenmodes with the highest values of Petermann factor (an other measure of non-normality) correspond to the eigenmodes with the highest projection coefficients $c_n$. Finally, we can conclude by saying that the optimal initial conditions (for a particular value of the propagation distance $z$) that lead to maximum transient growth are superposition of bound and radiation non-orthogonal modes with weights depending on their Petermann factor.

In the Appendix B, we find the optimal initial conditions and bounds for transient growth in two-dimensional (2D) potentials, demonstrating the generality of the findings discussed above. However, for 2D potentials with complex spatial distributions of gain and loss, it's computationally very demanding to compute the exact singular value decomposition (this requires the exponentiation of a large matrix), which is important for validation of our pseudo-spectrum estimates. To overcome this problem, we discuss an efficient computational method to truncate the propagator to a reasonable size, yielding evolution with controllable accuracy for a given $z$.

\section{Discussion and Conclusions}
\label{sec:end}

In summary, we present and study an unusual characteristic of a large class of non-normal photonic structures with distributed gain and loss but that are on average lossy. For wave propagation in such a medium, the common expectation is that the total optical power decays with increasing propagation distance. This is based on our notion of eigenvalues. If all eigenvalues are in the complex half-plane corresponding to decaying eigenmodes, any propagating beam will decay. In this paper we show that depending on the spatial distribution of the gain and loss there are optimal initial conditions for which the injected power can be amplified by several orders of magnitude, even though all the eigenmodes of the system are decaying. We systematically examine the characteristics (maximum amplification rate, transient growth dynamics and optimal initial conditions) of such non-normal power amplifiers in multimode photonic waveguides.

We note that our analysis based on pseudospectrum and singular value decomposition of non-normal operators is generic and can be directly applied to any non-normal optical system. In particular, our results  most likely have direct implications for the growing field of active plasmonics. To contrast our work to some earlier studies, coupled (\PT) symmetric plasmonic systems have been proposed for instance in Ref.~\cite{PT_plasmo} for optical switching in a directional coupler that employs one gainy and one lossy channel. Given a coupled system of gain and loss waveguides the gain that is required under single channel excitation (fixed initial conditions) for switching was examined. All the analysis was based on the existence of an exceptional point. To the contrary, we identify the initial conditions that lead to the maximum possible power amplification in any system that is dominantly lossy, without relying on the existence of any exceptional point. In our approach maximum growth is not due any switching from loss to gain regions, but a collective effect of all the modes (bound and radiation) of the total system. Additionally, in the most general case of an inhomogeneous gain-loss landscape (as the ones we consider), it can been shown that mapping of such potential to an equivalent (\PT) symmetric hamiltonian \cite{PT_exp} or a homogeneous lossy or gainy profile \cite{Siegman2} is mathematically impossible. This means that the physics and the methodology to understand the transient dynamics of lossy amplifiers is fundamentally different from both the (\PT)-passive optics\cite{PT_exp} and gain-guiding \cite{Siegman2}. Unlike such previous studies, our work applies the most general approach to understand and quantify the power growth in any dissipative system that has an arbitrary amount of gain, but still is overall lossy. In view of the above, we believe that the answers to our questions will have important implications to the efficient design of active plasmonic devices.

\section{Acknowledgements}
\label{sec:ack}

This work was in part supported by MIRTHE NSF EEC-0540832 and DARPA Grant No. N66001-11-1-4162. K. G. Makris is supported by the People Programme (Marie Curie Actions) of the European Union's Seventh Framework Programme (FP7/2007-2013) under REA grant agreement number PIOF-GA-2011- 303228 (project NOLACOME). L. Ge acknowledges partial support from PSC-CUNY 45 Research Grant.

\begin{appendix}

\section{Estimation of transient power growth in terms of pseudospectra \label{appA}}

Before we present the pseudo spectra analysis for the transient growth estimates, it is important and useful to first examine the spatial dimensions of our system. In particular, the normalizations for the spatial coordinates $x,z$, and the corresponding physical values for the gain-loss and the waveguide length are given in \cite{PT_theory}. As an example, for  $n_0=3.25$, $\lambda_0=1.55{\mu}m$,  ${\delta n} =0.001$, one normalized unit of $x$ corresponds to $2\lambda_0$,, while one normalized unit of propagation distance $z$ to $163\lambda_0$. This means that the waveguide dimensions are several times larger than the optical wavelength, and as a physical consequence the structure supports many guided modes. For the waveguide of Fig.~\ref{fig:schematics}(b), the spatial structure of these eigenmodes can be seen in the next figure (Fig.~\ref{fig:modes}). Specifically, the field amplitude of the 10 bounded modes of the non-normal potential (Fig.~\ref{fig:modes}(a-j)), their eigenvalues in the complex plane (Fig.~\ref{fig:modes}(k)) and their corresponding propagation dynamics (Fig.~\ref{fig:modes}(l)) are illustrated. Most of the spatial profiles of the modes are extended in both gain and loss regions and their eigenvalues (complex propagation constants) correspond to decay with the propagation distance $z$.

\begin{figure}[tb]
\includegraphics[clip,width=1.0\linewidth]{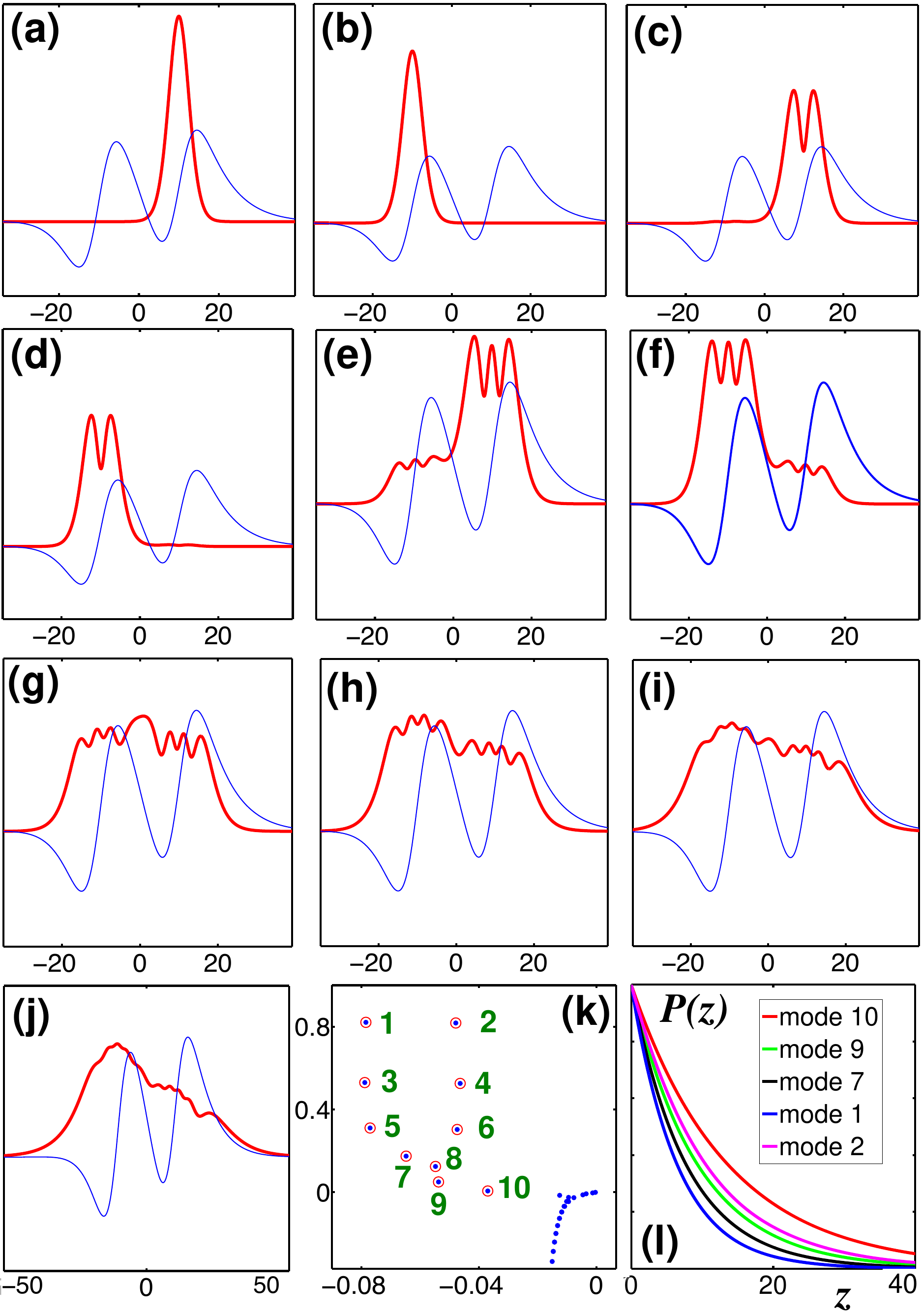}
\caption{(a-j) Spatial field amplitudes (normalized to 1) of the 10 bounded modes (red curves) of the non normal potential of Fig.~\ref{fig:schematics}(b), for $g=1$ as a function of the transverse coordinate $x$. The blue curves represent the imaginary part of the potential. (k) The eigenvalue spectrum in the complex plane. The red dots denote the ten guided modes of the structure. The eigenvalues are sorted based on the magnitude of their imaginary part. (l) power decay (normalized to 1) of the individually excited eigenmodes versus the propagation distance $z$.}\label{fig:modes}
\end{figure}

After understanding the properties of the eigenvalue spectrum, we are going to present how one can use its mathematical generalization for nonhermitian operators, the pseudo spectrum, in order to characterize the transient growth. The $\epsilon$ -pseudo spectrum of a non-normal operator $\Ahat$ is defined by the union of all its eigenvalues in the complex plane, when subjected to all possible system perturbations below a certain magnitude i.e. $\sigma_\epsilon(\Ahat) \equiv\{\mathfrak{z}\in\sigma(\Ahat+\Ehat), \forall\,\|\Ehat\|<\epsilon\}$. Here  $\sigma(\Ahat)$ denotes the spectrum of $\Ahat$. The norm of the operator $\Ahat$ is generally defined $\|\Ahat\|\equiv \sup_f\|\Ahat f\|/\|f\|$, where $\|f\|$ is the usual Euclidean norm of a function, $\|f\|^2\equiv\int_{-\infty}^{\infty} dx \, |f(x)|^2$. The parameter $\epsilon$ characterizes the strength of the random perturbations. Practically speaking, the pseudospectrum $\sigma_\epsilon(\Ahat)$ is a pattern in the complex plane that shows how sensitive the eigenvalue spectrum $\sigma(\Ahat)$ is to random perturbations. For Hermitian operators the spectrum and the pseudospectrum are almost identical. But for non-normal operators the two patterns can differ significantly, depending on the degree of the non-orthogonality of the corresponding eigenmodes. As seen in Fig.~\ref{fig:spa} showing the 0.1-pseudospectra for $g=0.5,1,2,3$, the closer we approach to the critical value $g_c=3.65$ for the emergence of positive eigenvalues of the unperturbed $\Ahat$, the more extended the pseudospectrum cloud becomes in the complex plane. This indicates an increased degree of non-normality as the angle between some of the eigenmodes becomes smaller, and the spectrum is more sensitive to random perturbations. By visual inspection of Fig.~\ref{fig:spa} one can immediately identify the modes that become more skewed and contribute more to the non-normal behavior.
The geometrical characteristics of the pseudospectrum cloud are directly related to the transient dynamics of paraxial equation of diffraction.  As such, we are interested to find order of magnitude estimates for the $G_{\max}$  in order to characterize our lossy amplifier.  In particular, the lower and upper bounds of $G_{\max}(z)$ can be estimated using theorems of functional analysis of non-normal operators \cite{TrefethenBook} and are directly associated with the extension of the pseudospectrum cloud to the right half complex plane for all $\epsilon\ll1$:

\begin{figure}[b]
\includegraphics[clip,width=1\linewidth]{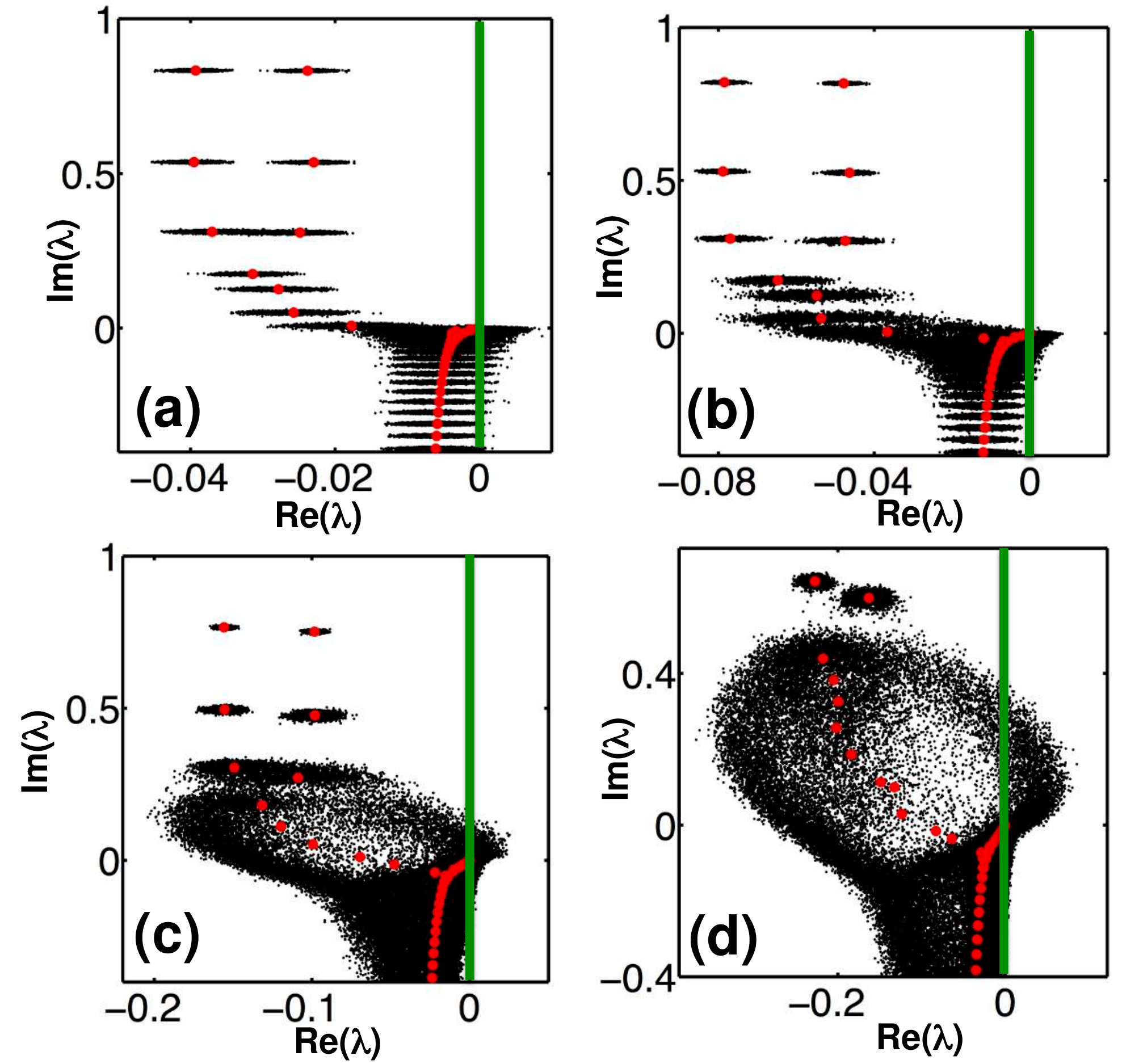}
\caption{Four $\epsilon=0.1$ pseudospectra of the non-normal potential shown in Fig.1, with the gain and loss amplitude $g$ is equal to (a) 0.5, (b) 1, (c) 2, and (d) 3. All plots are shown in the complex plane and the red dots depict the eigenvalue spectrum and the green line the imaginary axis.   } \label{fig:spa}
\end{figure}

\begin{figure}[tb]
\includegraphics[clip,width=1\linewidth]{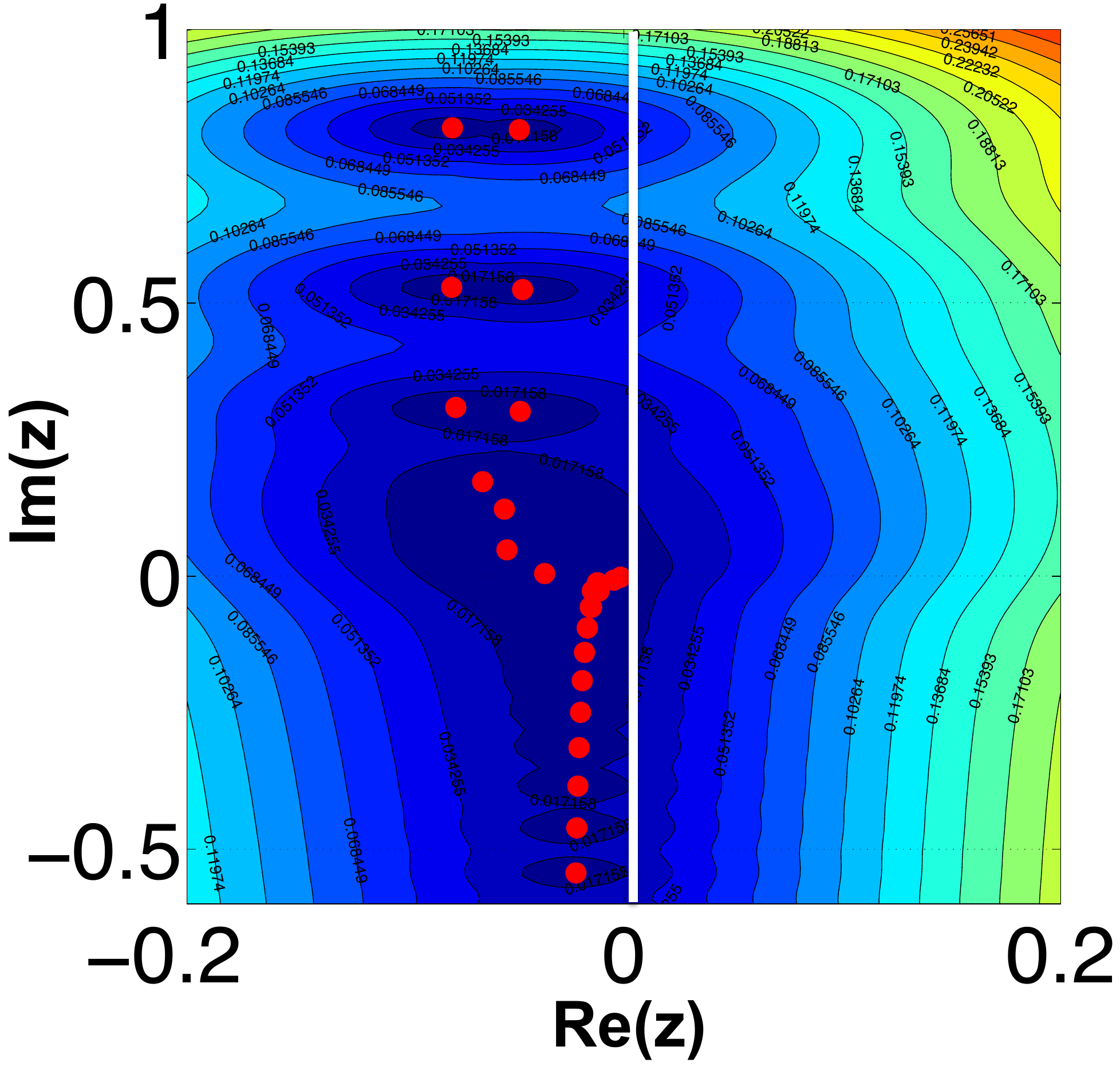}
\caption{Contour plots of different $\epsilon$-pseudospectra in the complex plane of the non-normal potential shown in Fig.1(b), for $g=1$. The red dots depict the eigenvalue spectrum.} \label{fig:con}
\end{figure}

\begin{align}
&\left\|e^{z\Ahat}\right\|\leq\frac{L_\epsilon\exp[z\alpha_\epsilon(\Ahat)]}{2\pi\epsilon}, \label{eq:upperBound}\\
&\left\|e^{z\Ahat}\right\|\geq \sup_\epsilon \frac{\alpha_\epsilon(\Ahat)}{\epsilon}. \label{eq:lowerBound}
\end{align}
Here the pseudospectral abscissa is defined as $\alpha_\epsilon(\Ahat)\equiv \max_{\mathfrak{z}\in\sigma_\epsilon(\Ahat)}\re{\mathfrak{z}}$
and is a measure of the extension of the pseudospectrum pattern into the upper half of the complex plane, whereas $L_\epsilon$ is a measure of the geometrical size of the whole pseudospectrum cloud. This approach is based in the Hille-Yoshida theorem, and provides the necessary and sufficient condition for transient growth \cite{growth1,growth2}. Namely, the pseudospectrum cloud must extend to the right half plane more than its $\epsilon = 0.1$ value.
Another way of accurate computation of pseudo spectra, is based on an alternative definition of the pseudo spectrum of a non normal operator in terms of the norm of its resolvent \cite{TrefethenBook}. In this case, the controur plot of Fig.~\ref{fig:con} describes many different pseudo spectra patterns in the complex plane for different values of $\epsilon$. By using both Fig.~\ref{fig:spa} and Fig.~\ref{fig:con} we can find order of magnitude estimates for the transient growth, based on the above two inequalities. For the specific potential that we are studying ($g=1$), we find that $L_{0.1}\sim0.8$ and $\alpha_\epsilon(\Ahat)\sim0.01$, leading to $\|e^{z\Ahat}\|\leq1.27\exp(0.01z)$. By calculating this upper bound for various values of the propagation distance $z$, we find the maximum of the upper bound of $G_{\max}(z)$ is reached roughly at $z=100$, with $\|e^{z\Ahat}\|^2\leq12$. We estimate also the lower bound for the transient growth for different values of $\epsilon$, and we find $\|e^{z\Ahat}\|^2\geq1.45$. Therefore our estimate for the maximum power growth for $g=1$ is $2\leq G_{\max} \leq 12$.
Therefore our estimation for the maximum power growth for the case of $g=1$ is $2\leq G_{\max}\leq12$.

\section{ Two-dimensional lossy potentials\label{appB}}

\begin{figure}[t]
\includegraphics[clip,width=1\linewidth]{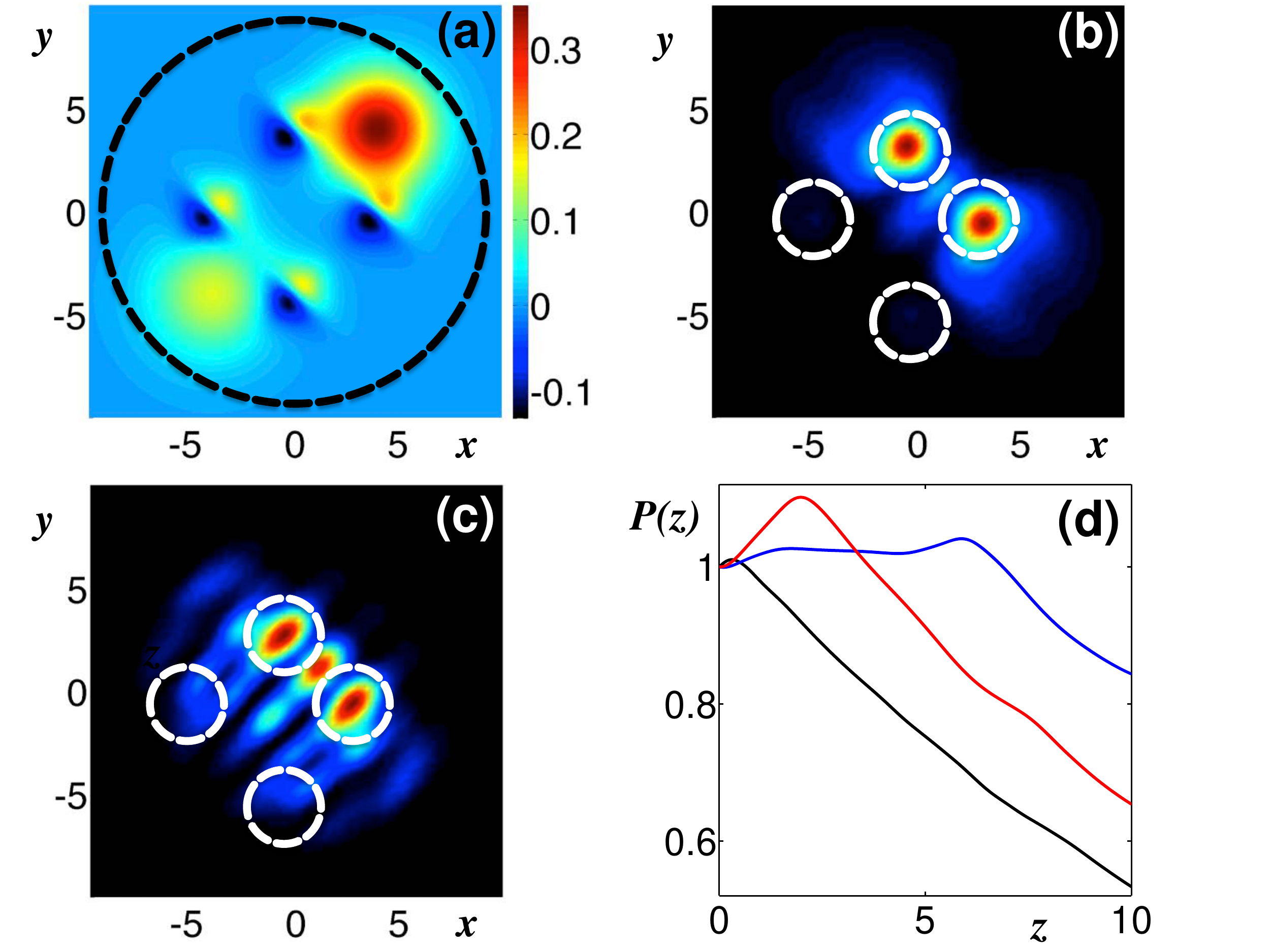}
\caption{(a) Spatial profile of a 2D non-normal optical potential in the transverse $x-y$ plane. The real part of the refractive index is a constant inside the dashed black curve and its imaginary part is given by the color scheme: positive values represent the lossy regions whereas the negative values represent the gain regions. (b,c) intensity of optimal initial conditions for achieving maximum transient growth at z=2,6, respectively. The dashed white circles represent the gain spatial regions. (d) Power dynamics versus the propagation distance $z$ for three different input waveforms. The black curve represents an intuitive waveform localized inside the gain regions. The red and blue curves represent the optimal waveforms shown in (b,c) that achieve the maximum growth at $z=2,6$, respectively.
}\label{fig:2D}
\end{figure}

In this section we extend our analysis to consider optical wave propagation in a non-normal two-dimensional (2D) potentials. In this case, the beam evolution is still governed by the two-dimensional paraxial equation of diffraction. This is a Schr\"{o}dinger-like equation $\frac{\partial \Phi}{\partial z}=\hat{H}\Phi$, with the evolution operator
\be
\hat{H} = i\ddx{} +i\ddy{} +iV(x,y).
\ee
We still require that the potential has net loss, meaning that the integral of $\im{V(x,y)} = gn_I(x,y)$ in the transverse plane is positive. A particular example of a multimode circular waveguide is depicted in Fig.~\ref{fig:2D} (a). Its real part is a constant inside the black dashed disk and it imaginary part by the color scheme. The methods of the previous section can be applied directly to this problem as well. Since it is computationally very intense and inaccurate to directly calculate the singular values and eigenvectors of a two-dimensional propagator, we use an alternative expansion method \cite{growth1,growth2} to calculate the transient growth $G_{\max}(z)$. This method is valid provided, that the eigenmode expansions are convergent and lead to accurate reconstruction of the projected field \cite{Siegman3}. In particular, by expanding the optical field to the biorthogonal eigenmode basis $\{\phi_n(x,y)\}$ and keeping only a finite number ($N$) of them, the field can be written as $\Phi(x,y,z) = \sum_{n=1}^N c_n \phi_n(x,y) e^{i\beta_nz}e^{\gamma_nz}$. One can show analytically that if the $\bm{F}$ is the Cholesky factorization of the hermitian matrix $\bm{B}_{mn} = { \inner{{\phi}_m}{{\phi}_n}  }$, then the transient power growth can be calculated by:
\be
G_{\max}(z) = \|\bm{F}\cdot\exp(z\bm{\Lambda})\cdot\bm{F}^{-1}\|^2,\label{eq:Gmax_2d}
\ee
where $\bm{\Lambda}$ is the diagonal matrix with the $N$ complex eigenvalues of the highest imaginary part. Here, instead of considering the propagator $\hat{G}$, we consider its finite approximation  $\hat{G}\approx\bm{F}\cdot\exp(z\bm{\Lambda})\cdot\bm{F}^{-1}$. In this way we calculate the optimal input waveforms that achieve the maximum growth at $z=2,6$ (Fig.~\ref{fig:2D}(b,c)). As we can see from these figures, their spatial patterns are far from the intuitive guess and do not localize in the gain regions, similar to what we found before for one-dimensional potentials. In Fig.~\ref{fig:2D}(d) we compare the power dynamics of these two input waveforms and of that localized only to the gain regions. Again the power growth for the latter is much less than what can be achieved using the optimal input waveforms determined by the right-singular vectors of $\bm{F}\cdot\exp(z\bm{\Lambda})\cdot\bm{F}^{-1}$ with the largest singular values.

\end{appendix}

\end{document}